# Faster, Higher, Stronger? The Impact of GenAI on Knowledge Work Productivity – Evidence from the Field

*Research Article*

SVEN BOTTESCH, University of Ulm
CHIARA SCHWENKE, University of Ulm
JAKOB ZIMMERMANN, Liebherr-Digital Development Center GmbH
MAXIMILIAN FÖRSTER, Karlsruhe Institute of Technology
MATHIAS KLIER, University of Ulm

**Abstract**

The rise of generative artificial intelligence (GenAI) has fueled high expectations regarding its potential to enhance knowledge work productivity in terms of efficiency and quality. Building on task-technology fit (TTF) theory, we empirically examine the extent of GenAI's productivity effect for different task types. We conducted a randomized lab-in-the-field experiment with 128 knowledge workers from a multinational industrial organization. Participants completed three representative knowledge work tasks (knowledge acquisition, packaging, and creation), either with or without GenAI. Results show that GenAI consistently increases efficiency across tasks. However, its impact on quality is task-contingent: quality increases for knowledge packaging and creation but declines for knowledge acquisition. Furthermore, GenAI tends to reduce quality variance for knowledge packaging and creation, primarily benefiting lower-performing knowledge workers. However, it increases quality variance for knowledge acquisition. These findings contribute to a more granular, differentiated understanding of GenAI's productivity impact and hold implications for research and practice alike.

## 1 Introduction

With the emergence of OpenAI's ChatGPT, Google's Gemini, Anthropic's Claude, and other tools, generative artificial intelligence (GenAI) has experienced a massive surge in practical use. Other than earlier forms of AI, GenAI does not produce output from training data in the form of decisions, but instead in the form of (seemingly) new content (Banh and Strobel 2023). GenAI is increasingly adopted in a variety of fields such as medicine (Thirunavukarasu et al. 2023), education (Kasneci et al. 2023), and business (Feuerriegel et al. 2024). Particularly in the latter, organizations have high expectations of productivity gains, e.g., by automating routine tasks, providing intelligent workflow assistance, and supporting innovation processes. GenAI's value potential for the global economy through increased productivity is estimated at 2.6 to 4.4 trillion US dollars, with the largest share being attributed to knowledge work (Chui et al. 2023). However, recent industry surveys have revealed that while many managers expect large productivity gains, GenAI support has mostly not (yet) materialized in financial gains (PricewaterhouseCoopers 2026). To more precisely characterize and eventually realize GenAI's value potential, a profound empirical account of its impact on knowledge work productivity is needed – both across the major types of knowledge work commonly encountered in organizational settings (knowledge acquisition, knowledge packaging, and knowledge creation) (Davenport et al. 1996) and across the major components of productivity (efficiency and output quality) (Tapasco-Alzate et al. 2022). However, existing research has not yet investigated all task types and both productivity dimensions within a single unified study design involving knowledge workers in their natural work environment. With this study, we aim to address this research gap and state the following research question:

**RQ:** *To what extent does the use of GenAI affect the productivity of knowledge workers for common knowledge work tasks?*

Task-technology-fit (TTF) theory proposes that productivity is contingent on the alignment of task requirements, individual abilities, and technological functionality (Goodhue and Thompson 1995). To address our RQ, we build on TTF theory to derive hypotheses on how the use of GenAI affects the productivity in knowledge acquisition, knowledge packaging, and knowledge creation. To test these hypotheses, we conducted a 2×3 mixed design lab-in-the-field experiment (Karahanna et al. 2018) in collaboration with a large multinational industrial organization. GenAI support served as the between-subjects factor with two treatment levels (with vs. without GenAI support), while task type (knowledge acquisition, knowledge packaging, and knowledge creation) constituted the within-subjects factor, as each participant completed all three tasks. A total of 128 knowledge workers from the organization were randomly assigned to two groups and asked to individually perform the three task types. Participants in the treatment group were supported by a GenAI application and participants in the control group had to complete the tasks without GenAI support. The experimental design allowed us to systematically assess the impact of GenAI support on efficiency (measured via task completion time) and output quality (measured via expert evaluations based on pre-defined constructs). We found that GenAI-supported knowledge workers completed tasks significantly faster across task types. The effect of GenAI on output quality was positive for knowledge packaging and knowledge creation, but negative for knowledge acquisition. Moreover, GenAI support was associated with a more stable performance, reflected in reduced variance across participants, which was mainly driven by lower-performing participants, who benefited more from GenAI support than higher-performing participants in terms of both efficiency and output quality. We expand existing research by quantifying the varying effect of GenAI on knowledge work efficiency and output quality across three distinct major types of knowledge work in a real-world organizational context. Our findings demonstrate the usefulness of TTF theory to anticipate GenAI effects on the productivity of knowledge work – at the same time, we theoretically extend TTF research by proposing a more nuanced view that distinguishes between the productivity dimensions efficiency and quality, as well as between different sub-dimensions of quality. Furthermore, our findings pave the way for IS researchers to develop and evaluate new GenAI approaches that are adaptive to task demands and employee skills. Organizations may use our findings to inform their GenAI strategies, fine-tune and individualize their respective organizational processes, and thereby avoid adverse effects.

The remainder of this paper is structured as follows: in the next two sections, we provide an overview of related work, followed by the theoretical background and development of hypotheses. Afterwards, we describe the methodological approach of this study, before we provide an overview of the results. Finally, we highlight our key findings and discuss implications for theory and practice. We conclude with a discussion of limitations and future research directions.

# 2 Related Work

Despite still being in its early stages of development, GenAI has triggered massive expectations regarding its impact on worker productivity. For example, Chui et al. (2023), in their scenario analysis based on expert inputs and existing research, estimate that GenAI could enable an annual global labor productivity growth of 0.1 to 0.6 percent. Notably, much of the existing research on (potential) productivity gains through GenAI is focused on knowledge work – which is understandable given the current dominance of text-generating models (Al Naqbi et al. 2024) and therefore also is the focus of our study. In the following, we provide a brief overview of knowledge work in general, introducing its definitions, task types, and productivity measurements. Furthermore, we offer a survey of recent literature on GenAI in knowledge work, synthesizing existing empirical findings on GenAI's productivity impact and identifying the research gap this study aims to address.

## 2.1 Knowledge Work

The term knowledge work emerged in the 1960s and 1970s, when researchers like Machlup (1962), Drucker (1969), and Bell (1973) anticipated the shift towards a post-industrial, knowledge-based economy – with the knowledge worker as the counterpart to the previously prevailing manual worker. Early definitions refer to the production of knowledge including its "disclosure, dissemination, transmission, and communication" (Machlup 1962, p. 7). Under the influence of technological, economic, and societal advancements, the concept of knowledge work has constantly evolved over time (Timonen and Paloheimo 2008). One way to differentiate the existing definitions is by their primary focus on knowledge as output of work (De Sordi et al. 2021), knowledge as resource or input of work (Iivari and Linger 2000), knowledge as a way of achieving the work (Kelloway and Barling 2000), or a combination of these (Newell et al. 2009).

To identify specific representative tasks of knowledge work, an activity-based perspective on knowledge work is appropriate (Kelloway and Barling 2000). Reich (1991) provides a foundational view by referring to the concept of symbolic-analytical services, which comprise non-standardized activities of problem solving, problem identification, and strategic brokering. Building on this perspective, Davenport et al. (1996) apply a process lens on knowledge work and, based on a study of 30 organizations, identify four primary activities: (1) knowledge acquisition, which entails understanding knowledge requirements, finding suitable elements in existing knowledge, and communicating them to the requester (e.g., research assistance); (2) knowledge packaging, which is concerned with the assemblage of knowledge that has been created independently (e.g., curation or editing); (3) knowledge creation, which entails generating new knowledge in its various forms (e.g., research or creative processes); (4) knowledge application, which refers to processes in which the creation of new knowledge is actively discouraged but the application of existing knowledge in itself is an intellectual task (e.g., in medicine). Out of those four, we focus on the initial three (knowledge acquisition, packaging, and creation) in this paper, as the latter (knowledge application) is of limited relevance in the organizational context we conducted our study in (see methodology section). Our focus is also in line with the most common uses of GenAI by knowledge workers to date beyond the context of our study (Brachman et al. 2024, 2025; Retkowsky et al. 2024).

Given the non-standardized nature of knowledge work, measuring its productivity does not only involve efficiency. Other than in Taylorism (Taylor 1911), which aims to identify the best way to perform a manual work task (as a standard) with the goal of realizing more units of production in a given timeframe, knowledge work productivity "is not […] primarily […] a matter of the quantity of output. Quality is at least as important" (Drucker 1999, p. 84). Accordingly, efficiency and output quality are the most frequently mentioned metrics in literature (Tapasco-Alzate et al. 2022), and we focus on these in our study as major dimensions of productivity. This is also in line with the suggestion of Dwivedi et al. (2023) that the impact of GenAI on knowledge work productivity should be studied with respect to efficiency and effectiveness (i.e., output quality). While efficiency can suitably be expressed by the time needed to fulfill a task (task completion time), an objective assessment of quality implies evaluating it against concrete, task-specific attributes and its conformance with stated requirements (Garvin 1988). It is therefore necessary to specify quality metrics for each knowledge work task in scope (which we do in the further course of this study, see methodology section).

## 2.2 GenAI in Knowledge Work

GenAI holds transformative potential for many occupations as it can reduce manual work by automating highly structured standard tasks and augmenting more complex tasks (Ooi et al. 2025; Söllner et al. 2024). This is particularly true in knowledge work, for which GenAI "offers new opportunities to recombine and leverage a company's knowledge assets" (Benbya et al. 2024, p. 24). Major use cases include (but are not limited to) artifact and idea creation, information search and analysis, and advice generation (Brachman et al. 2024). As Ulfsnes et al. (2024, p. 8) put it, knowledge workers are in "an unprecedented individual discovery and exploration phase", which can eventually lead to an augmentation of their productivity. Table 1 provides a structured overview of prior studies' empirical findings on GenAI's productivity impact in knowledge work.

**Table 1** Structured overview of previous studies on GenAI's productivity impact

| Paper | Task type(s) | Study design and sample | Main measurements of productivity | Key findings |
|---|---|---|---|---|
| Bouschery et al. (2023) | Knowledge creation (idea generation/creative brainstorming) | Lab experiment<br>168 university participants<br>GenAI tool: GPT-3 | Efficiency: number of ideas generated<br>Output quality: semantic prototypicality (proxy for creativity)<br>Algorithm-measured | ▪ Groups with GenAI support produce more ideas than groups without<br>▪ Ideas with GenAI support are also more creative than without<br>▪ AI-generated ideas drive the creativity advantage |
| Brynjolfsson et al. (2025) | Knowledge acquisition and packaging (customer support response writing) | Field experiment<br>5,172 customer-service agents from Fortune 500 firm<br>GenAI tool: GPT-3 | Efficiency: average handle time, resolutions/hour<br>Output quality: resolution rate, net promoter score<br>System-logged, customer-rated | ▪ GenAI-assisted workers resolve 15% more issues per hour overall, without significant changes in quality<br>▪ Performance gap between agents narrows, lower-skilled agents benefit most<br>▪ Skilled agents with small quality decline and only small efficiency gain |
| Chen and Chan (2024) | Knowledge creation (advertising copywriting) | Online experiment<br>355 panel-recruited (Prolific) participants, half with marketing experience<br>GenAI tool: GPT-4 | Efficiency: –<br>Output quality: ad clicks, perceived ad quality and creativity<br>System-logged, participant-rated | ▪ Outcomes differ across GenAI collaboration modes hinging on user expertise<br>▪ Using GenAI as a sounding board significantly improves quality for non-experts, closing the gap to expert content<br>▪ Using GenAI as a ghostwriter is significantly detrimental to experts' quality |
| Choi et al. (2024) | Knowledge creation (drafting legal documents, i.e., complaint, contract, memo, handbook) | Online experiment<br>59 students from a US law school<br>GenAI tool: GPT-4 | Efficiency: task completion time<br>Output quality: blind-graded quality scores<br>Self-reported, law professor-rated | ▪ GenAI significantly reduces task completion time across tested legal tasks<br>▪ Average quality improvement is significant only for contract drafting<br>▪ Where quality gains occur, they are concentrated in lowest-performing students, little to no benefit for top performers |
| Cui et al. (2026) | Knowledge creation (software development, coding) | Field experiment<br>4,867 software developers at Microsoft, Accenture, and a Fortune 100 company<br>GenAI tool: GitHub Copilot | Efficiency: number of weekly pull requests, commits, and builds<br>Output quality: build success rate (proxy for code quality)<br>System-logged | ▪ GenAI support significantly increases developer productivity by an average 26% (completed tasks)<br>▪ Code quality not significantly impacted<br>▪ Less experienced developers show significantly higher adoption rates and (directionally) higher productivity gains as compared to experienced developers |
| Dell'Acqua et al. (2026) | Knowledge packaging and creation (e.g., information synthesis, creative product ideation, business case analysis) | Lab-in-the-field experiment<br>758 management consultants from Boston Consulting Group<br>GenAI tool: GPT-4 | Efficiency: task completion time<br>Output quality: task completion rate, quality score, binary correctness, coherence quality<br>Expert-/algorithm-rated | ▪ GenAI increases productivity for tasks within its capability frontier: +12% task completion, 25% faster, higher quality<br>▪ For tasks outside GenAI's capability frontier, AI-assisted consultants are 19% less likely to produce correct output<br>▪ Lower-skilled employees benefit most on within-frontier tasks, i.e., performance gap narrows |
| Doshi and Hauser (2024) | Knowledge creation (creative writing of short stories) | Online experiment<br>293 non-professional writers, 600 non-professional reviewers, both panel-recruited (Prolific)<br>GenAI tool: GPT-4 | Efficiency: –<br>Output quality: novelty, usefulness, emotional characteristics<br>Evaluator-rated, self-assessed | ▪ GenAI increases individual creativity<br>▪ GenAI reduces collective novelty by rendering stories more similar to each other<br>▪ GenAI equalizes creativity scores between writers with low and high inherent creativity |
| Gambacorta et al. (2024) | Knowledge creation (software development, coding) | Field experiment<br>1,219 professional programmers from Ant Group<br>GenAI tool: CodeFuse | Efficiency: lines of code produced<br>Output quality: –<br>System-logged | ▪ GenAI increases the average code output by more than 50%<br>▪ Junior programmers (≤1 year experience) benefit most (+67%)<br>▪ Senior programmers without statistically significant gains, primarily due to lower engagement with the tool |
| Henkenjohann and Tenz (2024) | Knowledge creation (writing a blog post/hotel review) | Lab-in-the-field experiment<br>94 non-professional participants plus evaluators, both panel-recruited (Prolific)<br>GenAI tool: GPT-4 | Efficiency: –<br>Output quality: errors, structure, language, helpfulness, relevance, authenticity<br>Evaluator-rated | ▪ GenAI improves output quality regardless of collaboration pattern (delegation to or refinement by GenAI)<br>▪ However, writers' perceived responsibility for their work is reduced |
| Jia et al. (2024) | Knowledge creation (persuasive sales pitches) | Field experiment and qualitative interviews<br>40 professional sales agents serving 3,144 customers at a telemarketing company<br>GenAI tool: conversational bot (proprietary) | Efficiency: –<br>Output quality: customer questions outside training scope successfully answered (proxy for creativity), amount of sales<br>Algorithm-rated, manager-reviewed, system-logged | ▪ With GenAI support, sales agents' creativity more than doubles and sales are nearly twice as high than without<br>▪ Higher-skilled agents benefit more: top agents show substantially stronger creativity gains than bottom agents<br>▪ Higher-skilled agents develop positive emotions (mood/morale/freedom), while lower-skilled agents develop negative emotions (stress/defeat/lower morale) |
| Mehler and Krautter (2024) | Knowledge creation (software development, coding and debugging) | Online experiment<br>161 panel-recruited (Prolific) experienced Python developers<br>GenAI tool: GPT (no version specified) | Efficiency: –<br>Output quality: overall task performance, quality of generated code<br>Self-assessed, algorithm-rated | ▪ With GenAI support, algorithm-rated code quality improves for both the coding and the debugging task<br>▪ Self-assessed task performance improves only for coding (not debugging)<br>▪ Performance improvement for coding is mediated by reduced task difficulty |
| Noy and Zhang (2023) | Knowledge packaging (mid-level professional writing tasks, e.g., press releases or short reports) | Online experiment and follow-up surveys<br>453 college-educated professionals and evaluators, both panel-recruited (Prolific)<br>GenAI tool: GPT-3.5 | Efficiency: task completion time<br>Output quality: writing quality, content quality, originality<br>System-logged, evaluator-rated | ▪ GenAI reduces writing time by 40% and improves output quality by 18%<br>▪ Inequality between workers is reduced<br>▪ Workers with GenAI support remain significantly more likely to adopt GenAI in their real jobs after the experiment (2 weeks and 2 months after) |
| Peng et al. (2023) | Knowledge creation (software development, coding) | Online experiment<br>95 experienced, professional developers recruited via freelance platform (Upwork)<br>GenAI tool: GitHub Copilot | Efficiency: task completion time<br>Output quality: task success rate (standard automated GitHub Classroom test suite)<br>System-logged | ▪ With GenAI support, developers completed a standardized programming task 56% faster at a consistent quality level<br>▪ Developers with less experience and heavier daily coding load benefit most<br>▪ Also, relatively older age group (25-44) benefits more vs. younger developers |
| Przegalinska et al. (2025) | Knowledge creation (marketing tasks: product naming, competitive analysis, persona ideation, advertisement writing) | Lab experiment<br>94 graduate students and executive MBA candidates from European business school<br>GenAI tool: GPT-3.5 | Efficiency: –<br>Output quality: marketing-specific constructs (e.g., persona fit, brand awareness), linguistic metrics (e.g., readability)<br>Expert-/algorithm-rated | ▪ GenAI support improves participants' output quality across tasks<br>▪ (Prior) GenAI experience significantly positively impacts quality for 3/4 tasks<br>▪ GenAI produces output with more positive sentiment and simpler language than humans |
| Tsekouras et al. (2024) | Knowledge creation (writing of argumentative business essays) | Field experiment<br>2,305 undergraduate students from a European business school<br>GenAI tool: various (own choice) | Efficiency: –<br>Output quality: relevance, argumentation, evidence, writing quality<br>Evaluator-rated | ▪ GenAI improves writing quality but negatively impacts deeper cognitive processes (relevance, argumentation, evidence) initially (first essay)<br>▪ Negative impact diminishes over time (second essay), particularly for lower-performing students<br>▪ Performance gap between high- and low-performing students narrows |
| Vendraminelli et al. (2025) | Knowledge creation (writing of web article, conceptualization and execution) | Lab-in-the-field experiment<br>78 professionals with different occupations from global trading company<br>GenAI tool: web content tool (proprietary) | Efficiency: task completion time<br>Output quality: task-specific quality grades (incl., e.g., keyword relevance, formatting, grammar)<br>System-logged, evaluator-rated | ▪ GenAI overall fosters employee productivity with increases of quality and decreases of task completion time<br>▪ GenAI can reduce expertise gaps between rather similar occupations for conceptualization tasks<br>▪ For execution tasks and less similar occupations, a "GenAI wall" emerges (expertise gaps cannot be bridged) |
| Holzner et al. (2025) | Knowledge creation (various tasks) | Meta analysis across 28 studies with 8,214 participants | Efficiency: –<br>Output quality: creativity, diversity of ideas | ▪ GenAI significantly enhances individual creative performance but substantially reduces the diversity of ideas<br>▪ Laypeople benefit most from GenAI, academics and professionals show little to no creative performance gain<br>▪ Less creative individuals benefit more from GenAI than highly creative ones |
| Vaccaro et al. (2024) | Various types and tasks (grouped into creation and decision) | Meta analysis across 106 studies | Efficiency: –<br>Output quality: task performance (as per 370 different effect sizes) | ▪ Human-GenAI combinations on average underperform the best of either alone<br>▪ Losses in decision-making tasks but gains in content creation tasks<br>▪ Human-AI synergy only when humans outperform AI alone; when AI outperforms humans, combination with losses |

Existing empirical studies predominantly concentrate on knowledge creation tasks, such as software development (Cui et al. 2026; Gambacorta et al. 2024; Mehler and Krautter 2024; Peng et al. 2023), creative and professional writing (Doshi and Hauser 2024; Henkenjohann and Tenz 2024; Tsekouras et al. 2024; Vendraminelli et al. 2025), and ideation (Bouschery et al. 2023; Chen and Chan 2024; Przegalinska et al. 2025). In contrast, knowledge packaging is examined in only a small number of studies – e.g., Noy and Zhang (2023) and, partially, Dell'Acqua et al. (2026) – and knowledge acquisition has only been addressed by a single study yet (Brynjolfsson et al. 2025). Regarding the measurement of productivity, many studies omit the efficiency dimension. In those studies that measure it, GenAI support is consistently associated with reductions in task completion time or increases in output volume. For example, Peng et al. (2023) document that task completion time was more than halved for a standardized coding task with GenAI support; Brynjolfsson et al. (2025) report a 15% increase in customer issues resolved per hour; and Noy and Zhang (2023) report a 40% reduction in time needed to complete professional writing tasks. While its magnitude varies, the direction of the efficiency effect is broadly consistent. At the same time, evidence on output quality is more ambiguous. Quality improvements are reported across a range of knowledge creation and packaging tasks (Bouschery et al. 2023; Henkenjohann and Tenz 2024; Jia et al. 2024; Mehler and Krautter 2024; Noy and Zhang 2023; Przegalinska et al. 2025; Vendraminelli et al. 2025). However, quality effects are not uniform but rather conditional. A meta-analysis across 106 studies[1] (Vaccaro et al. 2024) finds that human-AI collaboration is associated with quality gains in creation tasks but quality losses in decision-making tasks. Another meta-analysis (Holzner et al. 2025) reports significant gains in individual creative performance but a substantial reduction in output diversity. At the level of individual studies, Dell'Acqua et al. (2026) observe that consultants' output quality with GenAI support increased for tasks within GenAI's current capability frontier but decreased for tasks outside that frontier. Choi et al. (2024) similarly find that quality gains are task-specific and that they are concentrated among low-performing participants. Chen and Chan (2024) show that output quality depends on how and by whom GenAI is employed: using GenAI as a sounding board significantly improves quality for non-expert users, whereas using it as a ghostwriter is detrimental to the quality of output produced by experts. The observation that lower-skilled, less experienced, or lower-performing individuals tend to benefit more from GenAI support emerges across other studies as well (Brynjolfsson et al. 2025; Holzner et al. 2025; Peng et al. 2023; Tsekouras et al. 2024). A notable exception is Jia et al. (2024), who report that higher-skilled sales agents experience greater performance gains when supported by a conversational GenAI tool.

This review of previous studies reveals three gaps in previous research, each of which our study aims to address. First, no prior research has examined all three major types of knowledge work (i.e., knowledge acquisition, packaging, and creation) within a single unified study. Existing work addresses the task types selectively, which makes it difficult to conduct systematic cross-task comparisons. Furthermore, knowledge acquisition has hardly been addressed in prior studies at all (except for Brynjolfsson et al. 2025). Unlike knowledge packaging and creation, knowledge acquisition tasks require precise and complete information retrieval, which often cannot reliably be achieved by GenAI (see theory section). The effects of GenAI on knowledge acquisition tasks might hence differ from those found for knowledge packaging and creation. Prior research has not addressed this question. The present study is, to the best of our knowledge, the first to investigate all three task types concurrently, using the same participants, GenAI tool, and evaluation framework. Second, the simultaneous assessment of both productivity dimensions (efficiency and quality) across multiple task types within a single study remains uncommon. Many studies measure only one dimension or capture both but focus on only one task type. This study's design, in which all participants complete tasks of all three major types and in which performance is assessed on both productivity dimensions (efficiency and quality), allows us to examine efficiency-quality relationships and task-type contingencies within a unified empirical framework. Third, the external validity of some of the existing evidence is limited due to constraints in sampling and measurement. While some field studies have been conducted, other studies rely on participant recruiting from crowdsourcing platforms such as Prolific or from student populations. While expert raters are involved in some studies, others assess productivity effects based on algorithmic quality proxies or self-reported performance measures. Our study addresses both concerns by conducting a lab-in-the-field experiment with knowledge workers employed in a multinational industrial organization and by having experienced evaluators from the same organizational context assessing output quality.

[1] It should be noted that the meta-analysis by Vaccaro et al. (2024) includes many studies beyond the scope of this research, i.e., studies that address AI in a wider sense and not GenAI in particular.

# 3 Theoretical Background and Development of Hypotheses

To examine how GenAI impact on knowledge worker productivity (potentially) varies across different tasks, we ground our research in task-technology fit (TTF) theory. TTF refers to "the correspondence between task requirements, individual abilities, and the functionality of the technology" (Goodhue and Thompson 1995, p. 218). When humans are supported by technology, their performance[2] improves if the capabilities of the technology align well with the requirements of the task (i.e., TTF is high). Building on this foundation, Tuczek et al. (2026) suggest that organizations conduct careful evaluations of task-GenAI-fit to guide appropriate strategies for GenAI deployment. Unlike earlier technologies, where decisions centered on adoption versus non-adoption, based on how structured and standardizable a task was, GenAI requires more nuanced choices on the continuum between full automation and selective augmentation of human work. Such choices are contingent on a richer understanding of task characteristics (Raisch and Krakowski 2021; Qian and Xie 2026). However, recent research shows that assessing

[2] In Goodhue and Thompson's (1995, p. 218) nomenclature, TTF leads to performance impact, which they define as "some mix of improved efficiency, improved effectiveness, and/or higher quality". This aligns with our two-dimensional definition of productivity comprising efficiency and output quality. We therefore use both wordings interchangeably in this paper.

TTF for GenAI is a non-trivial endeavor given the rapid evolution of GenAI and the still-developing understanding of its capabilities and limitations – thereby highlighting the need for further research that investigates GenAI from this theoretical lens (Dell'Acqua et al. 2026; Tuczek et al. 2026). Therefore, we base our research hypotheses on an assessment of GenAI's TTF for the three major knowledge work task types in scope for this study. Concretely, we discuss the characteristics of GenAI and contrast them with the task requirements of knowledge acquisition, knowledge packaging, and knowledge creation respectively. Following the logic that for high-TTF tasks the impact of GenAI on productivity is expected to be high as well (Goodhue and Thompson 1995), we then derive our research hypotheses.

Fundamentally, GenAI refers to a class of AI algorithms that produce (seemingly) new content, including but not limited to text. Modern GenAI is predominantly built on large-scale models pre-trained on vast corpora of data, so-called foundation models (Banh and Strobel 2023; Feuerriegel et al. 2024). Large language models (LLMs) are one type of foundation models for the language domain, containing millions to hundreds of billions (sometimes even more) of parameters. With increasing scale, LLMs demonstrate advanced capabilities, including in-context learning, instruction following, and multi-step reasoning, which make LLMs general-purpose models capable of solving a wide range of tasks (Wei et al. 2022; Zhao et al. 2026). However, LLMs are also subject to fundamental limitations, such as their restricted interpretability (black-box nature), the generation of plausible but factually wrong content (hallucination), or biases inherited from training data (Feuerriegel et al. 2024). In the following, we discuss those capabilities and limitations in more detail which are particularly relevant in the context of each the three knowledge work tasks.

First, knowledge acquisition entails understanding information needs, searching across multiple sources, synthesizing findings, and delivering the resulting information. Professional judgment is needed to ensure the output is accurate, reliable, and appropriate for the specific recipients and use context. The requirements of high-quality knowledge acquisition output hence primarily include aspects such as relevance, correctness, completeness, and timeliness (Davenport et al. 1996; Meyer and Zack 1996). Superficially, considering only the mechanical procedure of the task, GenAI seems well-suited given its broad pre-trained knowledge and near-instantaneous synthesis (Zhao et al. 2026). However, LLMs are designed to produce the statistically most probable next token given their training data, so that outputs are probabilistic approximations rather than guaranteed exact reproductions – which may result in hallucinations (or, reiterating the task requirements: in irrelevant, incorrect, incomplete, and non-timely output) (Brown et al. 2020; Ji et al. 2020). Retrieval-augmented generation architectures (RAG; as also used in this study) can help address this issue, as they allow for relevant content (that the model has not been trained on before, e.g., from a company database) to be fetched and incorporated at inference time, thereby grounding outputs in specific rather than general information (Gao et al. 2023; Lewis et al. 2020). Yet RAG architectures introduce their own sources of potential error that can compromise output accuracy, e.g., relevant documents may not be retrieved or may be dropped during context consolidation (Barnett et al. 2024), and even when relevant content is successfully retrieved, the model may be distracted by irrelevant passages, which can degrade factual correctness (Shi et al. 2023). Hence, there is a substantial mismatch: given the probabilistic nature of inference inherent to LLMs as well as the (potential) retrieval imprecision and sensitivity to irrelevant context inherent to RAG architectures, the task requirements remain difficult to be reliably achieved with GenAI. Taken together, these discrepancies between GenAI capabilities and knowledge acquisition task requirements suggest a low TTF. We therefore hypothesize:

**H1:** *GenAI support has a negative impact on average productivity of knowledge workers for a knowledge acquisition task.*

Second, knowledge packaging focuses on assembling, structuring, and presenting existing knowledge, which includes activities such as editing, proofing, summarizing, eliminating redundancy and contradiction, or format conversion. Professional judgment about how knowledge should be organized and presented is needed. The requirements of high-quality knowledge packaging output hence primarily include aspects such as completeness while focusing on salient content, coherence, and clear structure (Davenport et al. 1996; Hovy and Lin 1998; Over et al. 2007; Steinberger and Jezek 2009). GenAI is expected to be well-suited for this task due to the capabilities arising from its transformer architecture, i.e., the neural network design underlying modern LLMs. This architecture allows for the modeling of long-range dependencies and broader semantic patterns, which is relevant for the knowledge packaging task (Vaswani et al. 2017; Zhao et al. 2026). For example, the model can detect grammatical, stylistic, and content dependencies (and hence also incoherences) across extended text spans for editing and proofing. As LLMs are able to map semantically similar content regardless of surface phrasing, models can help identify equivalent and contradictory passages across documents for summarizing and redundancy elimination. And regarding format conversion, LLMs' advanced capabilities of instruction following and in-context learning allow a single model to produce structurally diverse outputs from identical source content simply by varying the prompt (Brown et al. 2020; Wei et al. 2022). Taken together, these overlaps of GenAI capabilities and knowledge packaging task requirements suggest a high TTF. We therefore hypothesize:

**H2:** *GenAI support has a positive impact on average productivity of knowledge workers for a knowledge packaging task.*

Third, knowledge creation refers to the task of generating novel and original knowledge, e.g., conducting research, writing original works, or developing strategies. It is less characterized by specific, reproducible sub-activities and more by its (literally) creative, autonomous nature. Outcomes are guided by intuition and accumulated experience. The requirements of high-quality knowledge creation output in a business context hence primarily include aspects such as the thoroughness of ideas, usefulness, rarity, and novelty (Davenport et al. 1996; Eisenberger and Selbst 1994; MacCrimmon and Wagner 1994; Plucker et al. 2004). GenAI's fit with this task is ambiguous. On the one hand, given LLMs' pre-training on vast amounts of human-generated text (i.e., knowledge), GenAI can rapidly produce numerous initial ideas, alternative framings, or hypotheses, providing a starting point for knowledge workers to elaborate. Given GenAI's capacity to generate fluent and coherent text through transformer-based architectures, it can transform rough idea notes into polished prose, leaving humans free to concentrate more on ideation (Feuerriegel et al. 2024; Zhao et al. 2026). On the other hand, LLMs are constrained in their capacity to generate genuinely novel content. By optimizing for statistically likely outputs, LLMs tend to gravitate toward consensus, conventional patterns, and average responses, limiting the unpredictability and serendipity of creative breakthroughs (Bender et al. 2021). Taken together, these (limited) overlaps of GenAI capabilities and knowledge creation task requirements suggest a moderate to high TTF. We therefore hypothesize:

**H3:** *GenAI support has a positive impact on average productivity of knowledge workers for a knowledge creation task.*

Finally, up to this point, we have focused on the impact of GenAI on the productivity of knowledge workers overall and have implicitly held individual characteristics constant. Yet individual characteristics, are common to vary and, as they represent the third component of TTF alongside task and technology characteristics, also shape task performance (Goodhue and Thompson 1995; Tuczek et al. 2026). Indeed, emerging empirical evidence indicates that low-performing knowledge workers benefit more from equivalent GenAI support on comparable tasks than their high-performing colleagues (Brynjolfsson et al. 2025; Peng et al. 2023; Tsekouras et al. 2024). These observed impact differences suggest varying levels of TTF for both groups, driven by individual characteristics. On the one hand, given their lower performance baseline, low-performing knowledge workers have more room for improvement (which GenAI can help them realize) than high-performing workers, who may therefore encounter a performance ceiling effect (Dell'Acqua et al. 2026; Vendraminelli et al. 2025). On the other hand, some early empirical evidence points to behavioral differences in human-GenAI interaction: high-performing knowledge workers tend to exhibit lower output acceptance but greater verification depth and selective judgment over AI outputs (Cui et al. 2026; Siu and Fok 2025), consistent with higher levels of domain knowledge and prior task expertise. Synthesizing these findings, we expect a higher TTF for low-performing than for high-performing knowledge workers. We therefore hypothesize:

**H4:** *GenAI support has a variance-reducing effect on output quality of knowledge workers across knowledge work tasks.*

# 4 Methodology

To test our hypotheses and investigate the impact of GenAI support on the efficiency and output quality of knowledge workers, we conducted a 2×3 mixed design lab-in-the-field experiment (Karahanna et al. 2018). GenAI support constituted a between-subjects factor: participants were randomly assigned to either a treatment group (with GenAI support) or a control group (without GenAI support). Knowledge work task type constituted a within-subjects factor: all participants completed three predefined knowledge work tasks (knowledge acquisition, knowledge packaging, and knowledge creation) in a fixed order. The study was carried out in collaboration with a large multinational industrial company. The study took place during the development phase of a GenAI application tailored for internal use, allowing our findings to directly inform its further deployment. The assessment of participant output was carried out by experienced knowledge workers and managers from the organization who routinely evaluate such work professionally.

## 4.1 Case Setting

Our study was conducted within a large, multinational, family-owned industrial group headquartered in Europe. Since its founding in the mid-20th century, the organization has expanded its operations across international markets and multiple industrial sectors, including advanced manufacturing, mechanical engineering, and automation technology. Today, it employs more than 50,000 people across 13 product segments and more than 40 independently managed production entities worldwide. This growth has resulted in various complexity challenges spanning, e.g., communication, onboarding, human resource inquiries, or data management. Therefore, as part of its digital transformation strategy, the organization initiated the development of an internal GenAI application. Unlike publicly available AI tools, which raise concerns around data privacy, regulatory compliance, and third-party dependence, this application was specifically designed to operate within the organization's secure IT infrastructure. During the development phase, we examined its impact on knowledge worker productivity.

## 4.2 GenAI Application as Treatment

The GenAI application served as the intervention for the treatment group in this study – not as the object of investigation itself. It was implemented using Microsoft Azure OpenAI, following a typical implementation pattern for integrating GenAI into existing organizational workflows (Microsoft 2025). The application relied on GPT-4o mini with a temperature setting of 0 to maximize reproducibility and comparability. To enable access to organization-specific knowledge, the application adopts a RAG architecture, which connects the language model to proprietary company data sources, allowing the model to generate responses grounded in accurate, up-to-date, organization-specific information (Gao et al. 2023; Lewis et al. 2020). The underlying database entailed information relevant to the tasks of the experiment. For vector embedding, the text-embedding-3-large model available through Azure OpenAI was used. For storage and retrieval Azure AI Search was employed – a cloud-based search service optimized for semantic and vector search. Relevance matching between user queries and stored content was performed using a combination of vector similarity search, key phrase search, and semantic reranking. For each query, the system retrieved the top five most relevant chunks ($k = 5$).

## 4.3 Representative Knowledge Work Tasks

The three representative knowledge work tasks our study examines (knowledge acquisition, packaging, and creation) were operationalized with the organization's leadership to reflect current issues of practical relevance to the company. In the knowledge acquisition task, participants were asked to extract two types of information from internal documents: qualitative information (specific research topics that had recently received investment) and quantitative information (differences in specifications of two current products). The task was presented in the form of a request from a supervisor preparing for a meeting with department heads and asking the participant to research and summarize the requested information in bullet points. In the knowledge packaging task, participants were asked to condense a text of 260 words (the description of a new product) into a summary of no more than four sentences. Continuing the storyline from the first task, the supervisor mentioned that the product would also be discussed during the mentioned meeting. In the knowledge creation task, participants were asked to generate innovative ideas for future product offerings and to suggest research priorities relevant for the company's long-term competitiveness. This task was again presented in the form of a supervisor request, this time inviting the participant to contribute creative input for an upcoming internal strategy workshop.

## 4.4 Experimental Procedure

The study was conducted between November 2024 and March 2025. Participants were randomly assigned to one of two experimental groups: a treatment group (hereafter referred to as w/ GenAI) and a control group (hereafter referred to as w/o GenAI). The only difference between groups was whether participants had access to the GenAI application during task completion or not. The study consisted of three phases: an introduction, a pre-survey, and the main study. The introduction was identical for both groups and included a standardized briefing on GenAI, covering its basic functionality and typical use cases (e.g., text generation using tools such as ChatGPT or Microsoft Copilot) to ensure a common baseline understanding across participants. In the pre-survey, participants were asked to provide their demographic information, educational background, job role, years of professional and leadership experience, and frequency of GenAI use in both professional and private contexts, to characterize the sample with respect to prior GenAI familiarity and to test for the effectiveness of random group assignment. In the main study, participants completed a knowledge acquisition, a knowledge packaging, and a knowledge creation task in a fixed order (potential implications of this design choice are discussed in the limitations). Each task followed a consistent three-step procedure: first, participants received a detailed task description. They were informed that they had received a request from their supervisor presented as a screenshot of an email. In this email, the supervisor addressed the knowledge worker with a request to perform the specific task. Second, participants worked on the task – either with the support of the GenAI application or without it, depending on group assignment. Participants in the treatment group had continuous access to the GenAI application, which was available in a separate browser tab. Since employees participated using their standard office device set-ups – which typically include two

screens – they were free to distribute the two browser tabs across screens, mirroring their actual day-to-day work setup. In contrast, participants in the control group completed the tasks without GenAI support. Both groups had manual access to the database of company-specific information (which was also underlying the RAG). Third, participants were asked to submit their responses as an e-mail answer directly below the task description on the same page. To ensure anonymity, they were instructed not to include their names in their responses. Figure 1 illustrates the task procedure for participants in the treatment group, including the task and submission screen (left) and the GenAI interface (right). Sensitive organizational information was blurred to ensure confidentiality.

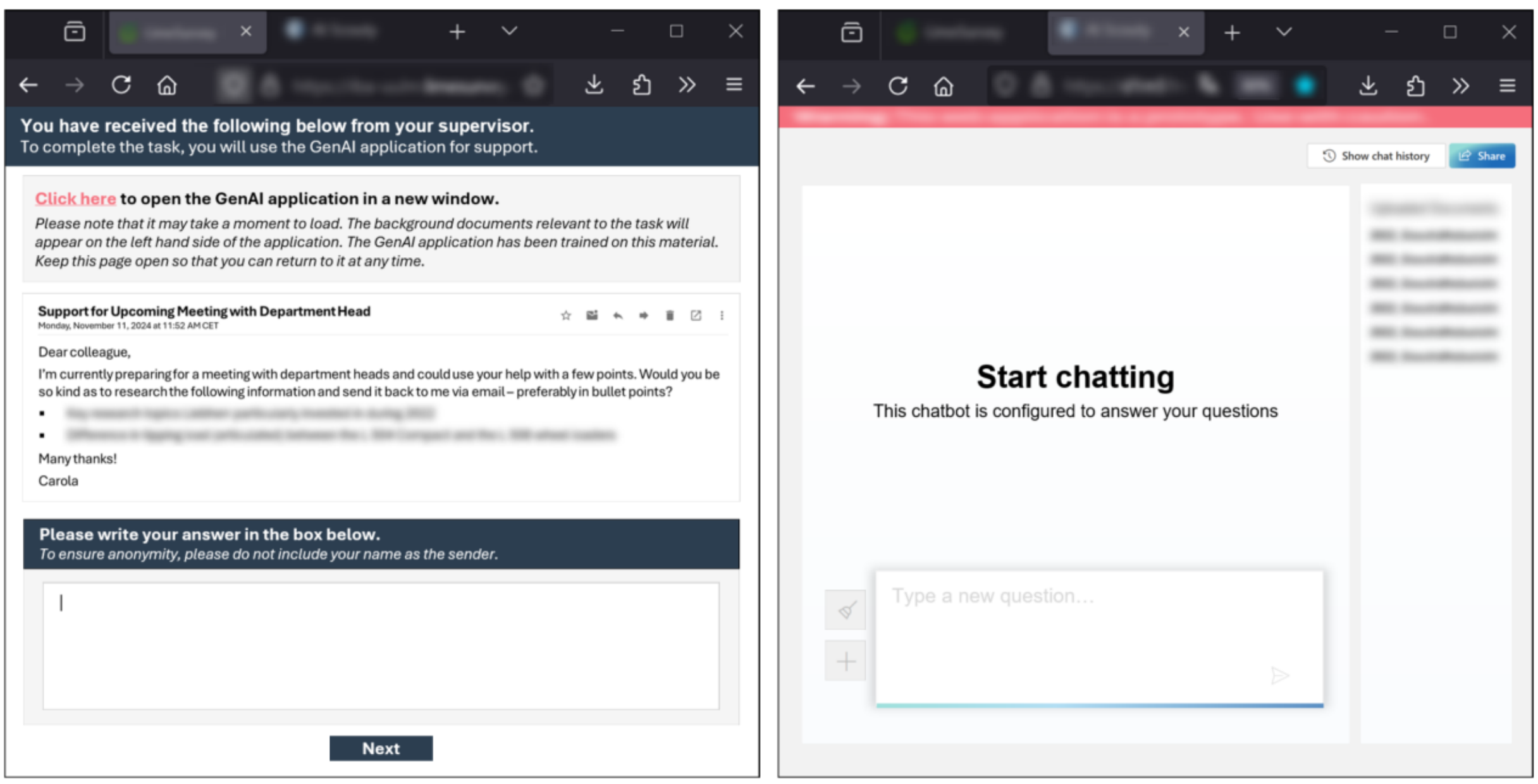

**Figure 1** Overview of the task procedure for the treatment group

## 4.5 Participant Sample

A total of 141 employees participated in the study. We excluded 13 incomplete responses: five participants did not commence the tasks after completing the pre-survey and were therefore not assigned to an experimental group, and eight did not complete all three tasks (five in the treatment group, three in the control group). The final sample hence comprised 128 participants, with 78 assigned to the treatment group (w/ GenAI) and 50 to the control group (w/o GenAI). The imbalance in group sizes is due to random assignment and the exclusion of incomplete responses, rather than any systematic allocation. The complete study took about 30 minutes per participant.

Of the final sample, 44 participants identified as female, 83 as male, and one preferred not to disclose their gender. The average age was 34.38 years (SD = 8.46). In terms of functional domains, 78 participants worked in technical roles (e.g., IT, product development, technical documentation) and 50 in non-technical roles (e.g., marketing, customer service, human resources). Participants across both technical and non-technical domains in the organization regularly encounter tasks involving knowledge acquisition, packaging, and creation as part of their professional responsibilities. Educational backgrounds ranged from secondary school level to doctoral degree, and the largest single group of participants reported more than 10 years of professional experience. Overall, the sample is broadly representative of the organization's workforce in terms of age, functional background, and professional experience. As shown in Table 2, participants varied considerably in their prior GenAI experience, ranging from never having tried GenAI to daily use, both professionally and privately.

**Table 2** Demographic characteristics of participants

| Characteristic | Full sample $n = 128$ | w/ GenAI $n_1 = 78$ | w/o GenAI $n_2 = 50$ |
|---|---|---|---|
| **Age: mean (SD)** | 34.38 (8.46) | 35.10 (9.30) | 33.26 (6.90) |
| **Gender** | | | |
| Female | 44 | 30 | 14 |
| Male | 83 | 47 | 36 |
| Not disclosed | 1 | 1 | 0 |
| **Functional domain** | | | |
| Technical | 78 | 48 | 30 |
| Non-technical | 50 | 30 | 20 |
| **Professional experience** | | | |
| Less than 1 year | 5 | 3 | 2 |
| 1-2 years | 13 | 7 | 6 |
| 3-4 years | 23 | 11 | 12 |
| 5-6 years | 18 | 9 | 9 |
| 7-8 years | 11 | 8 | 3 |
| 9-10 years | 6 | 4 | 2 |
| More than 10 years | 52 | 36 | 16 |
| **Leadership experience** | | | |
| Less than 1 year | 72 | 48 | 24 |
| 1-2 years | 16 | 6 | 10 |
| 3-4 years | 8 | 2 | 6 |
| 5-6 years | 15 | 10 | 5 |
| 7-8 years | 7 | 5 | 2 |
| 9-10 years | 3 | 2 | 1 |
| More than 10 years | 7 | 5 | 2 |
| **Education** | | | |
| Secondary school | 19 | 15 | 4 |
| Bachelor's degree | 50 | 29 | 21 |
| Master's degree | 52 | 30 | 22 |
| Doctoral degree | 7 | 4 | 3 |
| **Frequency of GenAI use (work-related)** | | | |
| Never tried | 4 | 1 | 3 |
| Tried only a few times | 29 | 25 | 4 |
| Monthly | 15 | 8 | 7 |
| Weekly | 26 | 14 | 12 |
| Daily | 54 | 30 | 24 |
| **Frequency of GenAI use (private)** | | | |
| Never tried | 9 | 8 | 1 |
| Tried only a few times | 32 | 19 | 13 |
| Monthly | 33 | 20 | 13 |
| Weekly | 35 | 20 | 15 |
| Daily | 19 | 11 | 8 |

To ensure the effectiveness of random group assignment before proceeding with data analysis, we tested for differences between the two experimental groups in socio-demographic variables and prior GenAI usage, as these variables could affect the impact of GenAI on knowledge work productivity. Mann-Whitney U tests for continuous variables and chi-squared tests for categorical variables yielded no significant differences between the two groups, indicating successful randomization.

### 4.6 Measurement and Data Analysis

The key dependent variable in our study is knowledge work productivity. In line with common definitions (see section on related work), we conceptualize productivity through two dimensions: efficiency and output quality. Efficiency was operationalized as task completion time. For each of the three tasks, time was recorded from the moment participants received the task description to the submission of their final response. This measure captured the full duration of task engagement, including interactions with the GenAI and potential double-checking in the information database (treatment group), manual search and review of information in the database (control group), as well as reflection, and preparation of a response to the supervisor (both groups). Task completion time was recorded in seconds.

Output quality was assessed based on submitted responses using a combination of task-specific and overarching, language-related evaluation criteria, grounded in prior literature and refined in collaboration with domain experts and senior managers from the organization. Given the task-dependent nature of quality in knowledge work (Garvin 1988), separate evaluation criteria were defined for each task. To operationalize these criteria, we derived an initial set of evaluation constructs from the literature, which was subsequently reviewed and refined in collaboration with domain experts and senior managers from the organization. This process ensured that the selected constructs were both theoretically grounded and appropriate for the specific organizational context (Dean et al. 2006).

For the knowledge acquisition task, the resulting evaluation criterion was *correctness*, which was defined as the degree to which the response accurately provides the requested qualitative and quantitative information. While prior literature conceptualizes knowledge acquisition quality along multiple dimensions such as relevance, correctness, completeness, and timeliness (Davenport et al. 1996; Meyer and Zack 1996), these aspects were collapsed into a single correctness construct in our setting. This is because the task required the extraction of highly specific, well-defined data points with a clearly defined ground truth, rendering the distinction between these dimensions conceptually negligible. Accordingly, responses were rated on a 3-point scale (1 = incorrect, 2 = partially correct, 3 = fully correct).

For the knowledge packaging task, output quality was assessed in terms of the attributes of high-quality summaries. These attributes can be derived from definitions in international standards and (computational) linguistics literature. The resulting evaluation criteria were *fidelity*, *representativeness*, and *conciseness*. *Fidelity* captures whether a summary sticks to information from the original text, synthesizes it accurately, and avoids adding any new information or opinions (adapted from DIN 1988; Hovy and Lin 1998; ISO 1976). *Representativeness* reflects whether a summary captures the central content of the original text without focusing on minor details or omitting any important aspects (adapted from DIN 1988; Over et al. 2007). *Conciseness* refers to whether a summary is shorter than the original source and non-redundant while remaining understandable and self-contained (adapted from ANSI 2015; DIN 1988; ISO 1976; Over et al. 2007).

For the knowledge creation task, output quality was assessed in terms of the attributes of high-quality (business) ideas. Following prior work in the creativity literature, the resulting evaluation criteria were *novelty*, *completeness, implicational explicitness*, and *relevance* (adapted from Dean et al. 2006). *Novelty* captures whether an idea is new and original, i.e., rare, inventive, or surprising. *Completeness* measures whether an idea is described concretely and in sufficient detail, avoiding vague statements. *Implicational explicitness* reflects whether an idea is well-justified regarding stated goals (in the context of our study superior customer value and long-term competitiveness, e.g., by referring to market potential, customer needs, or trends). *Relevance* captures whether an idea can be reasonably applied in an organizational context, e.g., by linking it to an existing portfolio or by proposing a meaningful extension.

In addition to these task-specific criteria, overarching, language-related evaluation criteria were applied to both the knowledge packaging and creation task. These included *clarity*, *coherence*, and *conventionality*. *Clarity* captures whether a text is written precisely and comprehensibly, e.g., by avoiding empty buzzwords, using specific and contextually clear language, and providing unambiguous references (adapted from ANSI 2015; DIN 1988; ISO 1976; Over et al. 2007). *Coherence* reflects whether a text has a logical structure and is well-organized, so that the readers can easily follow its logic (adapted from ANSI 2015; Hovy and Lin 1998; ISO 1976; Over et al. 2007). *Conventionality* refers to whether a text is written in a way that is appropriate for the target audience, using conventional wording and formatting as well as correct grammar and orthography (adapted from Hovy and Lin 1998; Over et al. 2007). All criteria for the knowledge packaging and creation tasks were rated on a 7-point Likert scale (1 = strongly disagree, 7 = strongly agree). Table 3 provides an overview of all output quality evaluation criteria.

**Table 3** Output quality evaluation criteria

| **Knowledge acquisition** | |
|---|---|
| **Construct** | **Task-specific evaluation criteria** |
| **Correctness** (qualitative information) | The response accurately provides the requested qualitative information. |
| **Correctness** (quantitative information) | The response accurately provides the requested quantitative information. |
| **Knowledge packaging** | |
| **Construct** | **Task-specific evaluation criteria** |
| **Fidelity** (DIN 1988; Hovy and Lin 1998; ISO 1976) | The summary sticks to information from the original text, synthesizes it accurately, and avoids adding any new information or opinions. |
| **Representativeness** (DIN 1988; Over et al. 2007) | The summary captures the central content of the original text (without focusing on minor details or omitting important points). |
| **Conciseness** (ANSI 2015; DIN 1988; ISO 1976; Over et al. 2007) | The summary is shorter than the original text and non-redundant – but still long enough to be understandable and self-contained. |

| **Knowledge creation** | |
|---|---|
| **Construct** | **Task-specific evaluation criteria** |
| **Novelty** (Dean et al. 2006) | The idea is new and original, i.e., rare, inventive, or surprising. |
| **Completeness** (Dean et al. 2006) | The idea is described concretely and in sufficient detail (i.e., not a vague statement like “leverage the potential of digitalization”). |
| **Implicational explicitness** (Dean et al. 2006) | The idea is well-justified regarding the goals of superior customer value and long-term competitiveness (e.g., does it address potential/market, customer needs, trends, etc.). |
| **Relevance** (Dean et al. 2006) | The idea can be reasonably applied in an organizational context (e.g., by linking it to the existing portfolio or by proposing a meaningful extension). |
| **Knowledge packaging and creation** | |
| **Construct** | **Overarching, language-related evaluation criteria** |
| **Clarity** (ANSI 2015; DIN 1988; ISO 1976; Over et al. 2007) | The text is written precisely and comprehensibly (e.g., avoids empty buzzwords, uses specific and contextually clear language, provides unambiguous references). |
| **Coherence** (ANSI 2015; Hovy and Lin 1998; ISO 1976; Over et al. 2007) | The text has a logical structure and is well organized, so that readers can easily follow its logic. |
| **Conventionality** (Hovy and Lin 1998; Over et al. 2007) | The text is written in an appropriate way for the target audience, using conventional wording and formatting as well as correct grammar and orthography. |

The quality assessment was carried out by four experienced knowledge workers and managers from the organization who routinely evaluate such work as part of their professional responsibilities. Each expert independently rated the quality of responses using the predefined criteria. To ensure the reliability of the assessment, we utilized the Intraclass Correlation Coefficient (ICC) employing the ICC(3,$k$) model. This model is based on a two-way mixed-effects design with absolute agreement for average ratings (Gisev et al. 2013; Koo and Li 2016; Lehmann et al. 2017). The calculated ICC values exceeded the recommended minimum threshold of 0.50 (Koo and Li 2016), indicating substantial agreement among the four expert raters. Specifically, for the knowledge packaging task, ICC values ranged from 0.74 to 0.88, and for the knowledge creation task, values ranged from 0.72 to 0.87 (see Table 4). These results justify the aggregation of scores, which is calculated as the average of the four ratings into a single quality measure for each participant and criterion. For the knowledge acquisition task, expert ratings were fully aligned, which is to be expected given the clearly defined ground truth.

**Table 4** Overview of measurement and data collection

| **Construct** | | **Scale** | **Data collection** | **ICC(3,*k*)** |
|---|---|---|---|---|
| ***Knowledge acquisition*** | ***Efficiency*** | | | |
| | Task completion time | Seconds (continuous) | System-logged (real-time timestamps) | – |
| | ***Output quality (task-specific criteria)*** | | | |
| | Correctness (qualitative information) | 3-point scale (1 = incorrect – 3 = fully correct) | Four independent expert raters (ground truth comparison) | – |
| | Correctness (quantitative information) | | | |
| **Knowledge packaging** | ***Efficiency*** | | | |
| | Task completion time | Seconds (continuous) | System-logged (real-time timestamps) | – |
| | ***Output quality (task-specific criteria)*** | | | |
| | Fidelity | 7-point Likert scale | Four independent expert raters | 0.76 |
| | Representativeness | | | 0.74 |
| | Conciseness | | | 0.85 |
| | ***Output quality (overarching, language-related criteria)*** | | | |
| | Clarity | 7-point Likert scale | Four independent expert raters | 0.74 |
| | Coherence | | | 0.75 |
| | Conventionality | | | 0.88 |
| **Knowledge creation** | ***Efficiency*** | | | |
| | Task completion time | Seconds (continuous) | System-logged (real-time timestamps) | – |
| | ***Output quality (task-specific criteria)*** | | | |
| | Novelty | 7-point Likert scale | Four independent expert raters | 0.83 |
| | Completeness | | | 0.84 |
| | Implicational explicitness | | | 0.87 |
| | Relevance | | | 0.72 |
| | ***Output quality (overarching, language-related criteria)*** | | | |
| | Clarity | 7-point Likert scale | Four independent expert raters | 0.87 |
| | Coherence | | | 0.82 |
| | Conventionality | | | 0.86 |

As our measures were not normally distributed (Shapiro-Wilk tests; $p<.05$), we conducted two-sided non-parametric Mann-Whitney U tests to compare the two experimental groups on each measure. This approach is robust to unequal sample sizes and does not rely on normality assumptions. To account for multiple comparisons across evaluation criteria, $p$-values were adjusted using the Bonferroni correction within each task domain (i.e., across all evaluation criteria associated with the same task). Specifically for testing H4, which hypothesizes a variance-reducing effect of GenAI support on output quality, we conducted Levene's tests for equality of variances between groups. Corresponding $p$-values were likewise adjusted using Bonferroni correction. To further explore potential heterogeneity in treatment effects beyond the main hypothesis tests, we conducted an exploratory top-bottom performer analysis. Participants were split into the top and bottom 50% within each experimental group for each quality evaluation criterion. This ensured sufficient group sizes while maintaining a clear distinction between higher- and lower-performing participants.

# 5 Results

This section presents the results addressing our hypotheses H1-H4. While the use of GenAI led to a significant reduction in the time required to complete all three tasks in the experiment, a more nuanced picture emerges when it comes to the quality of knowledge workers' output. The results are summarized in Table 5-7 and described for each task in the following. Post-hoc power analyses using G*Power 3 (Faul et al. 2007) suggest that the Mann-Whitney U tests achieved sufficient power ($1 - \beta \geq .76$).

**Table 5** Summary of efficiency and output quality analysis

| Construct | | w/ GenAI | w/o GenAI | H1 supported? | H2 supported? | H3 supported? |
|---|---|---|---|---|---|---|
| *Knowledge acquisition* | ***Efficiency*** | | | | | |
| | Task completion time | **989.43**$^{***}$ (611.48) | 1387.41 (733.38) | No | – | – |
| | ***Output quality (task-specific criteria)*** | | | | | |
| | Correctness (qualitative information) | 1.55 (0.62) | **1.86**$^{*}$ (0.61) | Yes | – | – |
| | Correctness (quantitative information) | 2.50 (0.86) | **2.88**$^{*}$ (0.48) | Yes | – | – |
| *Knowledge packaging* | ***Efficiency*** | | | | | |
| | Task completion time | **240.19**$^{***}$ (278.47) | 504.68 (342.37) | – | Yes | – |
| | ***Output quality (task-specific criteria)*** | | | | | |
| | Fidelity | **6.39**$^{*}$ (0.35) | 5.92 (0.91) | – | Yes | – |
| | Representativeness | **5.47**$^{***}$ (0.57) | 4.71 (1.07) | – | Yes | – |
| | Conciseness | **5.93**$^{***}$ (0.84) | 5.18 (1.22) | – | Yes | – |
| | ***Output quality (overarching, language-related criteria)*** | | | | | |
| | Clarity | 5.66 (0.66) | 5.33 (1.10) | – | No | – |
| | Coherence | 4.79 (0.75) | 5.11 (1.21) | – | No | – |
| | Conventionality | **6.39**$^{***}$ (0.56) | 5.38 (1.20) | – | Yes | – |
| *Knowledge creation* | ***Efficiency*** | | | | | |
| | Task completion time | **271.84**$^{***}$ (263.70) | 442.06 (256.34) | – | – | Yes |
| | ***Output quality (task-specific criteria)*** | | | | | |
| | Novelty | 3.59 (0.98) | 3.31 (1.20) | – | – | No |
| | Completeness | **4.54**$^{***}$ (0.99) | 3.43 (1.30) | – | – | Yes |
| | Implicational explicitness | **4.50**$^{***}$ (1.09) | 2.78 (1.28) | – | – | Yes |
| | Relevance | **6.09**$^{***}$ (0.35) | 5.61 (0.76) | – | – | Yes |
| | ***Output quality (overarching, language-related criteria)*** | | | | | |
| | Clarity | **6.12**$^{***}$ (0.50) | 5.15 (1.06) | – | – | Yes |
| | Coherence | **5.92**$^{***}$ (0.68) | 5.08 (1.06) | – | – | Yes |
| | Conventionality | **6.41**$^{***}$ (0.57) | 4.85 (1.08) | – | – | Yes |

***Notes:*** *Reported values are means with standard deviations. Statistical significance is based on two-sided Mann-Whitney U tests. To account for multiple comparisons, p-values were adjusted using Bonferroni correction within each task domain. Asterisks indicate statistically significant differences between the two experimental groups (with and without GenAI support) based on adjusted p-values (* $p < .05$, ** $p < .01$, *** $p < .001$).*

## 5.1 Impact of GenAI on the Productivity of Knowledge Acquisition

Regarding efficiency, participants with GenAI support completed the knowledge acquisition task significantly faster than participants without GenAI support, achieving an average reduction in completion time of 29%. In terms of quality, however, participants without GenAI support produced better results on average across both subtasks (see Figure 2). They gave higher-quality responses than participants with GenAI support for qualitative information (mean = 1.55 w/ GenAI vs. 1.86 w/o GenAI, $p$ = .010) and quantitative information (mean = 2.50 w/ GenAI vs. 2.88 w/o GenAI, $p$ = .010). Taken together, these findings partially support H1: while GenAI support improves efficiency (contrary to H1's prediction), it reduces output quality (consistent with H1's predicted negative effect on productivity).

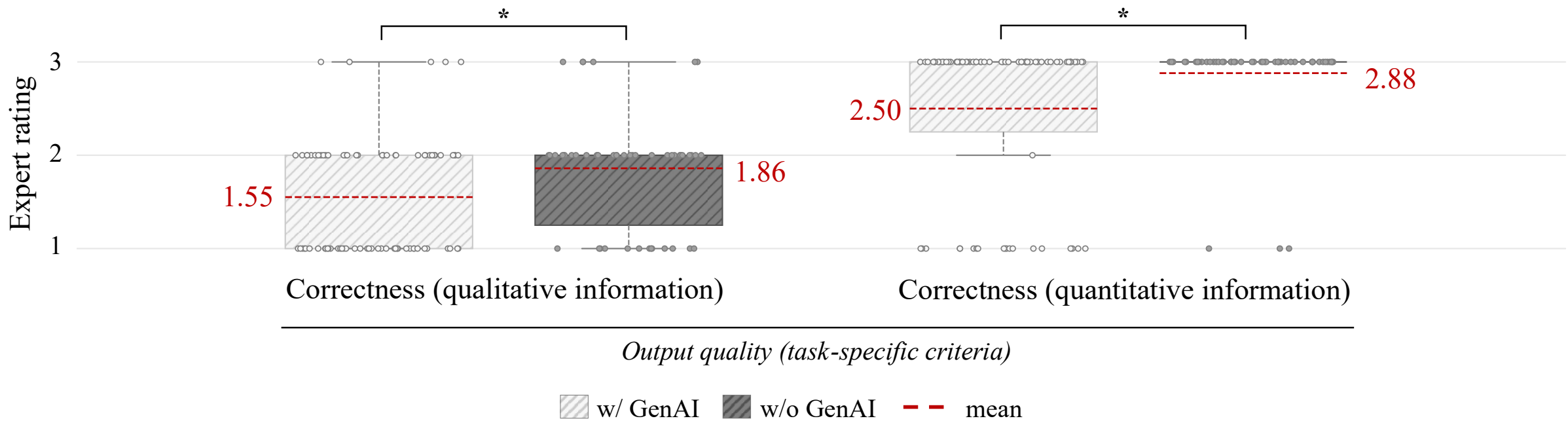

**Figure 2** Boxplots for the knowledge acquisition task

***Notes:*** *dots represent participants' mean evaluation scores; * p<.05, ** p<.01, *** p<.001 (according to Mann-Whitney U test; Bonferroni-corrected).*

## 5.2 Impact of GenAI on the Productivity of Knowledge Packaging

Regarding efficiency, participants with GenAI support completed the knowledge packaging task significantly faster than participants without GenAI support, achieving an average reduction in completion time of 52%. In terms of quality, participants with GenAI support outperformed those without support across four out of six evaluation criteria (see Figure 3).

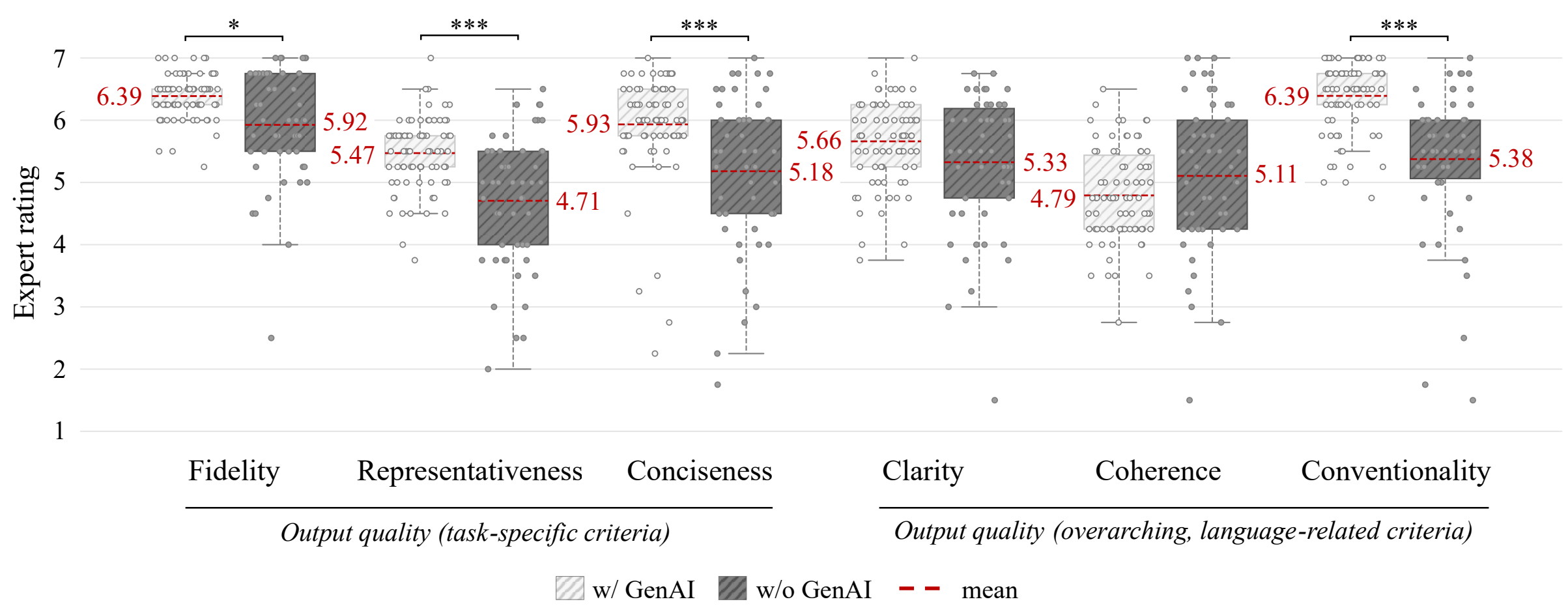

**Figure 3** Boxplots for the knowledge packaging task

***Notes:*** *dots represent participants' mean evaluation scores; * p<.05, ** p<.01, *** p<.001 (according to Mann-Whitney U test; Bonferroni-corrected).*

Statistically significant differences were observed for *fidelity*, *representativeness*, *conciseness*, and *conventionality*. The largest difference was found for *conventionality* (mean = 6.39 w/ GenAI vs. 5.38 w/o GenAI, $p < .001$). Taken together, these findings support H2: GenAI support improved efficiency and the output quality dimensions *fidelity*, *representativeness*, *conciseness*, and *conventionality* for knowledge packaging, despite no significant change in the language dimensions *clarity* and *coherence*.

## 5.3 Impact of GenAI on the Productivity of Knowledge Creation

As for the other two task types, the use of GenAI support also resulted in a reduction of completion time for the knowledge creation task with participants achieving an average decrease of 39% compared to those participants without GenAI support. In terms of quality, participants with GenAI support outperformed those without GenAI support across all seven evaluation criteria, on average. Statistically significant differences were observed for all criteria except *novelty* ($p = .471$). The largest difference between participants with and without GenAI support was observed for *implicational explicitness* (mean = 4.50 w/ GenAI vs. 2.78 w/o GenAI, $p < .001$). Taken together, these findings support H3: consistent with the hypothesized positive impact on average productivity, GenAI support improved efficiency and the output quality dimensions *completeness*, *implicational explicitness*, *relevance*, *clarity*, *coherence*, and *conventionality* for the knowledge creation task, despite no significant improvement in *novelty*.

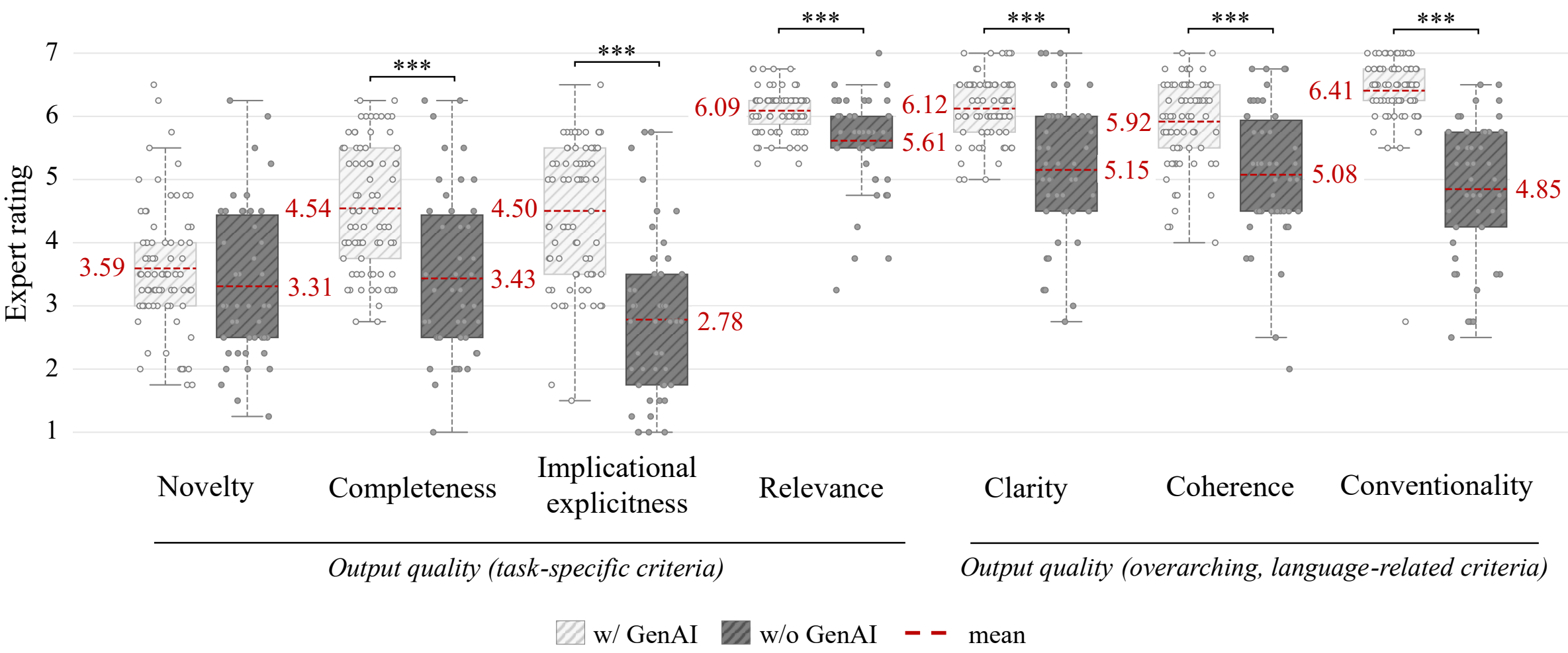

**Figure 4** Boxplots for the knowledge creation task

***Notes:*** *dots represent participants' mean evaluation scores; *p<.05, ** p<.01, *** p<.001 (according to Mann-Whitney U test; Bonferroni-corrected).*

## 5.4 Impact of GenAI on the Output Quality Variance across Knowledge Work Tasks

Regarding variance, the results differ notably across tasks (see Table 6 for key statistics and Table 7 for top-bottom performer analysis). For the knowledge acquisition task, no significant variance difference was observed for qualitative information ($p_{Levene} = .201$). For quantitative information, however, variance was significantly greater among participants with GenAI support ($p_{Levene} = .011$). As shown in Table 7, the performance gap between top and bottom performers was 1.00 in the GenAI group and 0.24 in the control group ($\Delta = +0.76$), indicating a substantially wider performance gap under GenAI support, driven by notably lower quality among bottom performers with GenAI support. These findings do not support H4 for knowledge acquisition. Contrary to the hypothesized variance-reducing effect, GenAI increases output quality variance. For the knowledge packaging task, significant differences were observed across all six evaluation criteria ($p_{\text{Levene}} < .050$), with consistently greater variance in the control group. This was primarily driven by the markedly lower quality of bottom performers without GenAI support compared to their counterparts with GenAI support. These findings support H4 for knowledge packaging. For the knowledge creation task, significant variance differences were observed for four out of seven criteria ($p_{\text{Levene}} < .050$), again with lower variance in the GenAI condition. Unlike the knowledge packaging task, top performers also benefited from GenAI support across several criteria: most notably *implicational explicitness* ($\text{mean}_{\text{top50}}$ = 5.43 w/ GenAI vs. 3.78 w/o GenAI) and *conventionality* ($\text{mean}_{\text{top50}}$ = 6.75 w/ GenAI vs. 5.73 w/o GenAI, see Table 7). Nevertheless, performance gaps between top and bottom performers were consistently narrower with GenAI support, as the stronger quality gains among lower-performing participants reduced overall variance. Taken together, these findings support H4 for knowledge packaging and knowledge creation, but not for knowledge acquisition.

**Table 6** Summary of variance analysis

| Construct | | $p_{\text{Levene}}$ | Variance direction | H4 supported? |
|---|---|---|---|---|
| ***Knowledge acquisition*** | ***Output quality (task-specific criteria)*** | | | |
| | Correctness (qualitative information) | 0.201 | – | No |
| | Correctness (quantitative information) | 0.011* | w/ GenAI > w/o GenAI | No |
| ***Knowledge packaging*** | ***Output quality (task-specific criteria)*** | | | |
| | Fidelity | < .001*** | w/o GenAI > w/ GenAI | Yes |
| | Representativeness | < .001*** | w/o GenAI > w/ GenAI | Yes |
| | Conciseness | 0.006** | w/o GenAI > w/ GenAI | Yes |
| | ***Output quality (overarching, language-related criteria)*** | | | |
| | Clarity | 0.005** | w/o GenAI > w/ GenAI | Yes |
| | Coherence | 0.002** | w/o GenAI > w/ GenAI | Yes |
| | Conventionality | 0.001** | w/o GenAI > w/ GenAI | Yes |
| ***Knowledge creation*** | ***Output quality (task-specific criteria)*** | | | |
| | Novelty | 0.193 | – | No |
| | Completeness | 0.401 | – | No |
| | Implicational explicitness | 1.000 | – | No |
| | Relevance | < .001*** | w/o GenAI > w/ GenAI | Yes |
| | ***Output quality (overarching, language-related criteria)*** | | | |
| | Clarity | < .001*** | w/o GenAI > w/ GenAI | Yes |
| | Coherence | 0.021* | w/o GenAI > w/ GenAI | Yes |
| | Conventionality | < .001*** | w/o GenAI > w/ GenAI | Yes |

***Notes:*** *Reported p-values are based on Levene's tests for equality of variances between the two experimental groups. The variance direction indicates which group exhibited greater dispersion in output quality. To account for multiple comparisons, p-values were adjusted using Bonferroni correction within each task domain. Asterisks indicate statistically significant differences in variance between the two experimental groups (with and without GenAI support) based on adjusted p-values (* p < .05, ** p < .01, *** p < .001).*

**Table 7** Summary of top and bottom performer analysis of output quality

| | Construct | Top 50% w/ GenAI | Top 50% w/o GenAI | Bottom 50% w/ GenAI | Bottom 50% w/o GenAI | Δ Gap |
|---|---|---|---|---|---|---|
| ***Knowledge acquisition*** | ***Output quality (task-specific criteria)*** | | | | | |
| | Correctness (qualitative information) | 2.10 (0.38) | 2.24 (0.44) | 1.00 (0.00) | **1.48*** (0.51) | +0.34 |
| | Correctness (quantitative information) | 3.00 (0.00) | 3.00 (0.00) | 2.00 (1.00) | **2.76** (0.66) | +0.76 |
| ***Knowledge packaging*** | ***Output quality (task-specific criteria)*** | | | | | |
| | Fidelity | 6.65 (0.19) | 6.61 (0.33) | **6.13*** (0.27) | 5.24 (0.77) | -0.85 |
| | Representativeness | **5.91** (0.30) | 5.55 (0.47) | **5.03*** (0.42) | 3.86 (0.78) | -0.81 |
| | Conciseness | **6.45* (0.27) | 6.12 (0.48) | **5.42*** (0.91) | 4.24 (0.98) | -0.85 |
| | ***Output quality (overarching, language-related criteria)*** | | | | | |
| | Clarity | 6.17 (0.29) | 6.16 (0.34) | **5.15** (0.52) | 4.49 (0.96) | -0.65 |
| | Coherence | 5.39 (0.48) | **6.08*** (0.60) | 4.19 (0.41) | 4.13 (0.81) | -0.75 |
| | Conventionality | **6.81*** (0.17) | 6.19 (0.42) | **5.97*** (0.51) | 4.56 (1.17) | -0.79 |
| ***Knowledge creation*** | ***Output quality (task-specific criteria)*** | | | | | |
| | Novelty | 4.30 (0.77) | 4.26 (0.92) | **2.88** (0.55) | 2.36 (0.45) | -0.48 |
| | Completeness | **5.39** (0.53) | 4.49 (0.96) | **3.68*** (0.44) | 2.38 (0.46) | -0.40 |
| | Implicational explicitness | **5.43*** (0.33) | 3.78 (0.97) | **3.59*** (0.72) | 1.78 (0.56) | -0.16 |
| | Relevance | **6.36*** (0.18) | 6.12 (0.30) | **5.83*** (0.27) | 5.11 (0.75) | -0.48 |
| | ***Output quality (overarching, language-related criteria)*** | | | | | |
| | Clarity | **6.53*** (0.24) | 6.01 (0.54) | **5.72*** (0.33) | 4.29 (0.69) | -0.91 |
| | Coherence | **6.45** (0.24) | 5.91 (0.56) | **5.39*** (0.54) | 4.24 (0.73) | -0.61 |
| | Conventionality | **6.75*** (0.20) | 5.73 (0.39) | **6.07*** (0.61) | 3.97 (0.80) | -1.08 |

***Notes:*** *Reported values are means with standard deviations. Participants were split into the top and bottom 50% based on their task-level quality score within each experimental group. Statistical significance is based on two-sided Mann-Whitney U tests. To account for multiple comparisons, p-values were adjusted using Bonferroni correction within each task domain. Asterisks indicate statistically significant differences between the two experimental groups (with and without GenAI support) based on adjusted p-values (* p < .05, ** p < .01, *** p < .001). Δ Gap reflects the change in performance gap between top and bottom 50% performers under GenAI support relative to the control group; negative values indicate a narrower gap under GenAI support, positive values indicate a wider gap.*

# 6 Discussion

The global rise of GenAI has fueled high expectations regarding its potential for productivity gains especially in knowledge work. Yet empirical evidence on the actual impact of GenAI in terms of both efficiency and output quality of knowledge work is still rare. Specifically, prior research has examined knowledge work task types selectively, rarely assessed efficiency and output quality simultaneously, and often relied on student populations or algorithmic quality proxies rather than experienced evaluators in real organizational contexts. Against this background, we hypothesized the impact of GenAI use on knowledge workers' productivity based on TTF theory and tested the hypotheses in a mixed design lab-in-the-field experiment with 128 knowledge workers from a multinational industrial organization to provide a more differentiated view of GenAI productivity effects.

Overall, this study offers three fundamental insights. First, GenAI support consistently improves efficiency across all three knowledge work task types. Second, the effect of GenAI on output quality is strongly task-contingent: while GenAI improves the majority of quality dimensions in knowledge packaging and knowledge creation, it reduces output quality in knowledge acquisition. Third, GenAI affects not only average performance, but also performance variance across workers. For knowledge packaging and knowledge creation, GenAI reduces variance in output quality, largely driven by stronger gains among lower-performing participants, whereas no such equalizing effect emerges for knowledge acquisition. Our findings contribute to research on GenAI in knowledge work by providing theoretically grounded empirical evidence on how its use affects productivity across all three major types of knowledge work. In the following, we discuss the implications of these findings for theory and practice.

## 6.1 GenAI Increases Knowledge Workers' Efficiency across Tasks

We found that GenAI support significantly enhances the efficiency dimension of productivity, measured as task completion time, across all three major task types of knowledge work. In line with TTF theory and our hypotheses H2 and H3, we found the largest time reduction for the high-TTF task knowledge packaging (-52%) and the second largest time reduction for the high-to-medium-TTF task knowledge creation (-39%). However, we also identified a substantial time reduction effect for knowledge acquisition (-29%). This finding is inconsistent with H1, in which we hypothesized an overall negative impact of GenAI support on productivity (efficiency and output quality) of knowledge acquisition due to its low TTF – i.e., although a low TTF proposes productivity losses (or at least no gains), we observed increases in efficiency.

The theoretical implication of this finding is that a more nuanced view on TTF is necessary in the context of GenAI, not assuming uniformly affected productivity but distinguishing between its dimensions. While it has been shown that TTF does not necessarily produce uniform effects across outcome types, e.g., perceptions, behaviors, and performance impacts (Howard and Rose 2019; Jeyaraj 2022), differential effects within one type (specifically performance) have been addressed to a limited extent. Future research might hence want to investigate under which circumstances outcome dimensions develop into opposite directions and which task characteristics particularly drive these differences.

Moreover, it is important to note that the unexpected reduction in task completion time for the knowledge acquisition task might be driven by users' tendency to overly rely on GenAI (Klingbeil et al. 2024), such that efficiency increases even in the absence of a strong TTF, potentially at the expense of quality. Indeed, we observed a deterioration of the quality dimension of productivity for knowledge acquisition (as will be discussed in Section 6.2). Organizations implementing GenAI may hence face the risk of faster completion masking reductions in output quality. It is therefore imperative for researchers and practitioners alike to develop systems of human-GenAI collaboration that are carefully calibrated based on task characteristics. Design choices and deployment decisions should be based on explicit cost and benefit analyses that weigh potential efficiency gains against potential quality losses. As our study shows, TTF generally provides a suitable lens to guide respective further IS research and design.

## 6.2 GenAI Enhances Output Quality in some Tasks – and Reduces It in Others

We found task-contingent effects of GenAI support on output quality of knowledge work. In line with TTF theory and our hypotheses H1-H3, our findings show that GenAI enhances overall output quality for tasks with high TTF, while reducing it for tasks with low TTF. Taken together, our findings demonstrate the usefulness of TTF theory to anticipate GenAI effects on the quality of knowledge work. At the same time, the absence of significant effects across all quality criteria suggests that even within the quality dimension (and not only between the two productivity dimensions efficiency and output quality, as discussed in Section 6.1) a more nuanced view of TTF is required to account for variation in the magnitude and consistency of improvements. In the following, we discuss these nuances.

For knowledge packaging, GenAI support led to clear improvements across task-specific quality criteria: knowledge workers synthesized the original text more accurately (*fidelity*), captured the central aspects of the original text more precisely (*representativeness*), and found a better balance between text length and understandability (*conciseness*). As hypothesized, the capabilities of LLMs, such as modeling long-range dependencies and following structured instructions, appear to align closely with the core requirements of knowledge packaging tasks, enabling knowledge workers to produce better outputs – at least regarding task-specific quality criteria. Notably for the overarching, language-related quality criteria, GenAI support did improve conventional wording, grammar, and orthography (*conventionality*) but it did not have a significant effect on precise and comprehensible writing (*clarity*) or logic structure (*coherence*). These findings extend previous research which has found that GenAI support helps improve writing quality (e.g., Noy and Zhang 2023). For a task like knowledge packaging, in which the synthesis of text represents the core nature of the task, to which knowledge workers can attribute their entire mental capacity during fulfillment, GenAI might have little to add in terms of language-related quality.

This is different for the knowledge creation task, in which the core nature of the task is creative brainstorming and textual synthesis of ideas rather represents a finalizing step. We found significant improvements across all language-related criteria for the knowledge creation task. In terms of the task-specific quality criteria, GenAI also helped knowledge workers produce more concrete (*completeness*), goal-fitting (*implicational explicitness*), and context-specific (*relevance*) knowledge (in this case business ideas). However, participants did not create business ideas that were more original, rare, inventive, or surprising (*novelty*) when supported by GenAI. This seems to reflect a fundamental limitation of LLMs, namely that by optimizing for statistically probable outputs, these models tend to gravitate toward conventional patterns, limiting their capacity to stimulate genuinely original thinking (Bender et al. 2021). Our results extend earlier findings from, e.g., Doshi and Hauser (2024), who observed that for general tasks, which are not dependent on organization-internal data and are hence supported by standard-architecture GenAI systems, collective novelty was reduced as outputs were rendered more similar to each other. Our findings show that this relationship also holds for more specific tasks, which are dependent on organization-internal data and are hence supported by RAG-architecture GenAI systems. By grounding outputs in the index corpus, GenAI output appears to anchor participants to existing organizational knowledge, thereby constraining rather than expanding the novelty of their ideas.

For knowledge acquisition, we found clear evidence that GenAI support reduces output quality, consistent with H1, given the low TTF of this task type: the probabilistic nature of LLM inference is fundamentally misaligned with the precision and correctness requirements of knowledge acquisition tasks. The *correctness* of both quantitative and qualitative information to be acquired was significantly lower with GenAI support. For example, when asked to find and extract quantitative information, approximately 25% of responses in the treatment group either misreported or fabricated numerical values, even though the correct information was readily available in the underlying database. Given that, other than hypothesized, knowledge workers needed less time for the completion of the knowledge acquisition task with GenAI support, we argue that the concurrent quality decline is very likely driven by overreliance (Buçinca et al. 2021; Klingbeil et al. 2024). Rather than using GenAI output as a starting point for verification – which would have required additional time and a switch of cognitive processes from search to validation – participants appear to have accepted it uncritically – not investing additional time and not switching cognitive processes. Such behavior would be in line with earlier evidence from Lee et al. (2025), who found that when tasks are perceived as routine or low-stakes, knowledge workers tend to forgo critical scrutiny of GenAI output, assuming GenAI is competent for tasks deemed trivial.

These results have important implications for research and practice. Preliminary practitioner-level findings from a multi-method research study on GenAI adoption across industries suggest that many organizations are still in experimentation phase, exploring general-purpose tools, without measurable (financial) productivity impact. The (fewer) firms that already realize productivity gains from GenAI support tend to run GenAI systems that are adapted to specific workflows, retain context, and improve over time (Challapally et al. 2025). Seen in concurrence with these observations and the continuing exponential development of GenAI, our findings reinforce the call for differentiated, organization- and task-specific, dynamically evolving GenAI strategies to realize the technology's productivity potential. The theoretical framework of task-GenAI tensions and fit proposed by Tuczek et al. (2026) can serve as a great starting point. We hope that our study will encourage further IS research into this direction. For practitioners, our findings can provide initial guidance for strategic managerial action. The results of our study corroborate the proposition of Challapally et al. (2025) that GenAI support needs to be integrated properly with workflows. This includes task-specific system designs (as discussed in Section 6.1) but also education and constant development of knowledge workers to avoid overreliance on GenAI, for example by drawing on approaches from XAI (Bauer et al. 2021). Future research could design and evaluate mechanisms such as confidence scores, detailed traceable citations, or uncertainty cues to support knowledge workers in calibrating their reliance on GenAI (Schneider 2024).

## 6.3 GenAI Reduces the Output Quality Gap for some Tasks – and Widens It for Others

We found the effect of GenAI on output quality variance to differ markedly across task types. Other than hypothesized (H4), GenAI reduces variance for knowledge packaging and knowledge creation but increases it for knowledge acquisition. This pattern suggests that the variance-reducing effect of GenAI is contingent on TTF. For high-TTF tasks, GenAI appears to provide a stabilizing effect that reduces quality differences between knowledge workers, whereas for low-TTF tasks, GenAI amplifies quality differences.

To better understand the drivers of this stabilization effect, we conducted an exploratory analysis comparing the top and bottom 50% of performers. These results should be interpreted with caution given their exploratory nature. With this caveat in mind, the findings suggest that the reduction in variance is primarily driven by improvements among lower-performing participants, while higher-performing participants show comparatively smaller gains. Our study hence confirms prior research which found that lower-performing individuals tend to benefit more from GenAI support than their higher-performing counterparts (Brynjolfsson et al. 2025; Choi et al. 2024; Peng et al. 2023; Tsekouras et al. 2024). However, our study also extends prior research by showing that the output quality gap is not reduced for all tasks. Where it occurs, one possible explanation for the variance reduction lies in the tendency of LLMs to generate statistically probable outputs. This leads to more homogeneous output content of GenAI-supported knowledge workers (Bottesch 2025; Dell'Acqua et al. 2026; Doshi and Hauser 2024) and hence might also be responsible for output quality homogenization. As knowledge workers become increasingly experienced in collaborating with GenAI and develop more individualized interaction strategies, this homogenization effect could potentially attenuate over time, representing a relevant avenue for future longitudinal research.

Furthermore, prior research suggests that while AI can augment lower-skilled employees, it may simultaneously challenge the role identity of higher-skilled workers, as their distinctive contributions become less visible (Strich et al. 2021). In consideration of this issue, our findings highlight the need for future research to design GenAI systems that not only elevate lower-performing users but also actively support and amplify the capabilities of high-performing ones (Abdel-Karim et al. 2023; Dellermann 2019). For organizations, the potential of GenAI to reduce performance disparities and raise the overall quality baseline may seem promising. However, it may blur traditional markers of expertise and reduce differentiation among employees. It may also conceal the need for lower-performing (particularly junior) knowledge workers to perform less sophisticated work unsupported on their own before being able to advance to higher levels of skill and work sophistication successfully (Puranam 2024). Organizations will therefore need to reconsider how expertise is defined and developed in work environments increasingly shaped by GenAI (Salamito et al. 2024). Such considerations should also inform the formulation of comprehensive GenAI strategies, which we discussed in Section 6.2.

# 7 Limitations and Future Research

While our study has revealed interesting initial insights into the impact of GenAI on knowledge work productivity, it is subject to several limitations, which at the same time offer opportunities for future research. First, the study was conducted within a single organizational context. While this setting enhances the real-world relevance of our findings, they may not readily generalize to other types of organizations, such as small and medium-sized enterprises, or to environments where GenAI is already embedded in routine workflows. In addition, our study follows a lab-in-the-field design, combining a real-world organizational setting with a controlled task environment. While this strengthens ecological validity, the standardized nature of the tasks may limit generalizability to fully naturalistic work settings. Future research should examine how GenAI applications perform across a broader range of organizational types and work settings to develop an even more differentiated understanding. Second, our study was conducted in early 2025 and provides timely insights in a nascent field. But given the fast and continuous evolution of GenAI and its underlying technology, it would be beneficial to observe GenAI-supported knowledge workers in their everyday work over extended periods. Longitudinal studies would be valuable to assess long-term effects of GenAI on knowledge work productivity and how these effects change over time. Third, our study was conducted during the development of a specific GenAI application for use within the collaborating organization. While the application reflects state-of-the-art practices and is representative of many enterprise-grade solutions, it remains a single implementation. Future research could investigate other model classes and explore alternative model configurations (e.g., differently adjusted parameters in LLMs). Fourth, our experimental design focused on three representative knowledge work tasks performed independently. While this allowed us to isolate task-specific effects, it does not capture how efficiency and quality unfold in interdependent or sequential task structures common in real-world contexts. Moreover, although the task sequence reflects a realistic workflow, the fixed order may introduce spillover effects that cannot be fully ruled out. Future research could examine continuous workflows combining knowledge acquisition, packaging, creation, and application, where efficiency gains in one task may influence subsequent tasks, for instance through error or bias propagation requiring rework. Finally, our study has an exploratory character and, grounded in TTF theory, provides a further step toward understanding the impact of GenAI on knowledge work productivity. While we show that GenAI can significantly influence efficiency and output quality across tasks, we do not directly examine the underlying mechanisms. In addition, our explorative analysis of how knowledge workers at different performance levels are impacted is based on a median split derived from expert ratings rather than independently measured skill levels. Future research could draw on psychological and behavioral theories to uncover these mechanisms, employ independent measures of expertise to enable more rigorous analyses of performance heterogeneity, and leverage interaction logs or process-tracing approaches to better understand how usage patterns shape GenAI-supported knowledge work. By addressing these limitations, future work can deepen our understanding of GenAI's role in knowledge work and help refine its design, deployment, and evaluation.